\documentclass{aa}
\usepackage{graphicx}

\def\kms{$\rm \,km\,s^{-1}$}

\def\Hubble{$\rm \,km\,s^{-1}\,Mpc^{-1}$}
\def\H2{H$_2$}
\def\Bg{Br$_{\gamma}$}
\def\NGC1068{NGC\,1068}
\def\micro{$\rm \,\mu m$}
\def\densflux{$\rm \,erg\,s^{-1}\,cm^{-2}\,\mu m^{-1}$}
\def\fluxsr{$\rm \,erg\,s^{-1}\,cm^{-2}\,sr^{-1}$}
\def\flux{$\rm \,erg\,s^{-1}\,cm^{-2}$}
\def\power{$\rm \,erg\,s^{-1}$}
\def\pc{\,pc}
\def\etal{et al.}

\def\Xray{X-ray}
\def\Xrays{X-rays}
\def\north{North}
\def\south{South}
\def\east{East}
\def\west{West}
\begin{document}  

\title{Near-IR 2D-Spectroscopy of the 4''$\rm \times$4'' region
around the Active Galactic Nucleus of NGC\,1068 with ISAAC/VLT  
\thanks{ Based on observations collected at the ESO/Paranal 8m UT1 telescope, 
Proposals 63.P-0167A \& 66.B-0142A}} 
\author{ E.\,Galliano\and D.\,Alloin  
 }
\institute{ European Southern Observatory, Casilla 19001, 
Santiago, Chile
}
\offprints{E.Galliano} 
\mail{egallian@eso.org}
\date{Submitted to Astronomy \& Astrophysics}
\titlerunning{Near-IR 2D-Spectroscopy of NGC\,1068}

 
\abstract{
New near-IR long slit spectroscopic data obtained with ISAAC on VLT/ANTU 
(ESO/Paranal) complement and extend our previously published near-IR  
data (Alloin \etal~\cite{all01}) to produce \Bg~and \H2~emission line maps 
and line profile grids of the central 4''$\times$4'' region surrounding the 
central engine of \NGC1068. The seeing quality together with the use of an 
$0.3$'' wide slit and $0.3$'' slit position offsets allow one to perform 
2D-spectroscopy at a spatial resolution $\approx 0.5$''. Slit orientations  
(PA=102\degr~and PA=12\degr) were chosen so as to match respectively the 
equatorial plane and the axis of the suspected molecular/dusty torus in 
\NGC1068. The selected wavelength range from 2.1 to 2.2\micro~is suitable 
to detect and analyze the \Bg~and \H2~emission lines at a spectral 
resolution corresponding to 35\kms. An asymmetric distribution of \H2~emission around 
the continuum peak is observed. No \H2~emission is detected at the location of
the strong 2.2\micro~continuum core (coincident within error-bars with the central engine 
location), while two conspicuous knots of \H2~emission are detected at about 1'' 
on each side of the central engine along PA=90\degr, with a projected 
velocity difference of 140\kms: this velocity jump has been interpreted in Alloin \etal~(\cite{all01}) as the 
signature of a rotating disk of molecular material. From this new
data set, we find that only very low intensity \Bg~emission is detected at the location of the 
two main knots of \H2~emission. Another knot with both \H2~and \Bg~emission 
is detected to the North of the central engine, close to the radio source C where the 
small scale radio jet is redirected and close to the brightest [OIII] 
cloud NLR-B. It has a counterpart to the South, placed almost symmetrically 
with 
respect to the central engine, although mainly visible in the \Bg~emission.
The northern and southern knots appear to be related to the ionization cone.  
At the achieved spectral resolution, the \H2~emission line profiles appear 
highly asymmetric with their low velocity wing being systematically more 
extended than their high velocity wing. A simple way to account for 
the changes of the \H2~line profiles (peak-shift with respect to the 
systemic velocity, width, asymmetry) over the entire 4''$\times$4'' 
region, is to consider that a radial outflow is superimposed over the emission 
of the rotating molecular disk. We present a model of such a kinematical
configuration and compare our predicted \H2~emission profiles to the observed 
ones. Excitation of the \H2~line is briefly discussed: X-ray irradiation 
from the central engine is found to be the most likely source of excitation. Given the fact that the material obscuring our direct view toward the central engine is Compton thick ($\rm N_{H} \geq 10^{24}\,cm^{-2}$), the observed location of the main \H2~knots at a distance of 70 pc from the central engine suggests that the rotating molecular disk is warped.     
\keywords{Galaxies : \NGC1068
--~Galaxies : Seyfert
--~Galaxies : nuclei
--~Galaxies : molecular gas
--~Galaxies : active
--~Infrared : galaxies 
--~Instrumentation : Near-IR}                    
}

\maketitle


\section{Introduction}
Over the past decade, considerable efforts has been devoted to a better 
understanding of active galactic nuclei (AGN). In particular, the use of 
HST (UV and visible ranges), the use of adaptive optics (AO) on 4m-class 
telescopes (near-IR range) and the use of interferometry (millimeter 
and radio ranges) have brought a large gain in spatial resolution and
hence many new constraints for AGN modeling. Conversely, the so-called 
unified model (see Krolik \cite{kro99}) has played an important role in 
stimulating the search for AGN constituents: in particular for the 
elusive pc-scale molecular/dusty torus which is thought to surround 
the central engine (black hole and accretion disk) and to funnel the 
ionizing radiation within a cone. Models of the infrared (IR) emission 
of the torus published so far, have explored torus radii from 1 
to 100\pc~(Krolik \& Begelman \cite{kro86}, Pier \& Krolik \cite{pie93}, Granato \& Danese \cite{gra94},  
Efstathiou \& Rowan-Robinson \cite{efs94}, Granato, Danese \& Franceschini \cite{gra97}), 
while the thickness, composition and inclination of the torus are 
other key-parameters in the modeling. Given the input energy distribution 
from the central engine, models then predict the flux spatial distribution 
at IR wavelengths from 2.2\micro~up to 20\micro~in the central 100\,\pc~region 
of the AGN. If one chooses an AGN 
for which the spatial resolution 
reachable today (typically 0.1'' with the facilities mentioned 
above in a range from UV to millimetric) provides an intrinsic scale 
relevant to the torus size, a comparison between predicted IR maps 
and observed ones will be an excellent opportunity for deriving 
stringent constraints on the model parameters. This calls for 
investigating a close-by and torus-inclined AGN: \NGC1068 is an 
ideal target.

This galaxy is located at a distance of 14.4\,Mpc, assuming a Hubble 
constant of 75\Hubble. The corresponding scale is 70\pc~per arcsec. 
Many high resolution images have indeed been obtained over the past decade:
decisive contributions come from the HST in the UV, visible 
(Capetti \etal~\cite{cap97}) \& near-IR (Thompson \etal~\cite{tho01}), 
from AO in the near-IR (Marco \etal~\cite{mar97}, Thatte \etal~\cite{tha97}, Rouan \etal~\cite{rou98} and Marco 
\& Alloin \cite{mar00}), using shift-and-add techniques or diffraction-limited 
observations on 8m class telescopes in the mid-IR (Bock \etal~\cite{boc00}, 
Tomono \etal~\cite{tom01}), from millimeter interferometry 
(Helfer \& Blitz \cite{hel95}, 
Tacconi \etal~\cite{tac97}, Backer \cite{bac00}, Schinnerer \etal~\cite{sch00})
 and radio interferometry 
(Gallimore \etal~\cite{gal97}). In this paper, we consider the compact radio 
source S1 (nomenclature after Gallimore \etal~\cite{gal96}) as the nuclear 
reference, that is the central engine location. Several observational 
facts suggest the presence of a molecular/dusty torus 
in \NGC1068. The term torus is used here in the sense of a rotating disk-like 
distribution. From inner to outer scales, let us 
point out the most relevant ones:
\begin{itemize}
\item
The 1\pc-size disk of ionized material seen almost edge-on at 
PA=110\degr, within S1, which has been detected at 8.4\,GHz with the VLBA: 
it is thought to trace the inner walls of the torus 
(Gallimore \etal~\cite{gal97})
\item
The core and the 15\pc-size elongated structure (at PA=102\degr), detected 
at 2.2\micro~with CFHT AO/PUEO (Rouan \etal~\cite{rou98}) and also present on 
3.5 and 4.5\micro~images with ESO AO/Adonis (Marco \& Alloin \cite{mar00}); the 
position of the near-IR core is found by Marco \etal~(\cite{mar97}) to be 
coincident with S1 within 
$\pm 0.05$''. This 
configuration is consistent with thermal emission from dust particles at 
1500K in the core, and 600K at a radius of 15\pc~and could trace the body 
of the torus.
\item
From interferometer maps (Backer~\cite{bac00}, Schinnerer \etal~\cite{sch00}) one observes, in addition to a yet 
unresolved core of low level CO emission, two prominent CO emitting peaks located 
symmetrically 1'' (70\pc) away from S1 at PA=98\degr~and separated in velocity 
by 100\kms: the authors interpret the CO observed emission in terms of a warped CO disk.
\end{itemize}

The possible presence of a molecular/dusty torus could be unveiled through the 
analysis of the \H2~roto-vibrational $\nu$=1-0\,S(1)~emission line at rest 
wavelength 2.122\micro. Indeed, excitation of molecular transitions such
as \H2~2-1\,S(1) and \H2~1-0\,S(1) can occur from irradiation by UV photons or 
X-rays, as 
well as from shocks, all ingredients which are found in profusion in an AGN. In addition,
as the \H2~emission arises from warm molecular gas, the study of the \H2~lines 
provides physical information complementary to that derived from CO molecular 
transitions. 

Molecular hydrogen was detected for the first time in an extragalactic 
source in the central region of \NGC1068 by Thompson, Lebofsky \& Rieke (\cite{tho78}). 
They also detected \Bg~emission in this region. Further \H2~line observations 
were carried out by Hall \etal~(\cite{hal81}), and Oliva \& Moorwood 
(\cite{oli90}). The first attempt at resolving spatially the \H2~emitting 
region was made by Rotaciuc \etal~(\cite{rot91}): from their best-seeing map 
they found that the \H2~line emission was very weak at the location of the 
central engine and was mainly arising from two knots of unequal 
intensity located at $\sim$1.3'' from the nucleus along PA=70\degr. A second 
attempt to get a $\leq 1$'' resolution image in the \H2 line emission was 
made by Blietz \etal~(\cite{bli94}). In the latter study, in contrary to 
Rotaciuc \etal~(\cite{rot91}), the \H2 emission was found to arise in three
knots, the brightest one being coincident (within an error bar of $\pm$0.3'') 
with the strong near-IR continuum core (central engine location).

In this study, we analyze the \H2~and \Bg~emission lines, at respective 
rest wavelengths 2.122 and 2.166\micro~using ISAAC on VLT/ANTU, in its short 
wavelength medium resolution mode. Preliminary results about the 
\H2~emission within a small region (1.5''$\times$1.5'') around the central 
engine were already published in a Letter (Alloin \etal~\cite{all01}). In the
current paper we extend the size of the region explored to 4''$\times$4'' around the nucleus and provide  \H2~line profiles and \H2~line emission 
maps over this region. We also analyze, 
similarly to \H2, the \Bg~line emission map and line profiles.
From the \H2~line profile analysis, we build a kinematical model of the
warm molecular component in the AGN of \NGC1068.
    
We present in Section 2 the relevant information for data collection 
and data processing. In Section 3 we provide complete sets of the \H2~and \Bg~line profiles across the entire region explored.
From these data sets, emission maps in the \H2~and \Bg~lines have been reconstructed and are compared to the map in the near-IR 
continuum. Fluxes and velocities are also given in Section 3. 
From the \H2~line peak velocities and profiles, we discuss in Section 4 the 
kinematics of the molecular material and we present in Section 5 a simple kinematical model, aiming at reproducing the profile shapes. Possible sources of excitation of
the \H2~line in the close environment of the AGN in \NGC1068 are discussed
in Section 5. Our concluding remarks are presented in Section 6.
\section{Data: Collection and Reduction} 

\subsection{Instrumental setup}
\begin{center}
\begin{figure}[h]
\mbox{\resizebox{8cm}{8cm}{\includegraphics[scale=1.]
{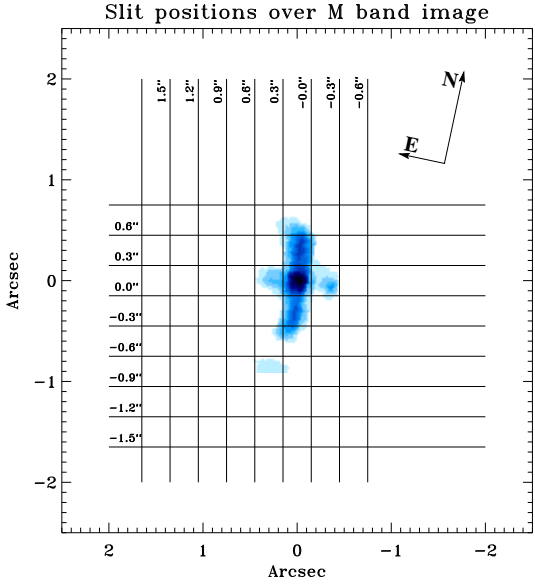}}}

\caption{Slit positions superimposed over the M band image from Marco \& Alloin (\cite{mar00}). Horizontal figures 
are along PA=102\degr~while  vertical ones are along PA=12\degr.}
\label{ref_positions}
\end{figure}  
\end{center}

Observations were performed using the SWS1 short wavelength arm of 
ISAAC, in medium resolution mode (MR), attached to VLT/ANTU.
Due to the request for very good seeing quality, observations were performed 
in service mode. Results discussed in this paper are based 
upon data acquired in July and August 1999, December 2000, and January and 
February 2001. A summary of 
the observations is given in Table~\ref{obs_summary}.
\begin{table}
\caption[]{Summary of the observations}
\begin{center}
\begin{tabular}{cccc} \hline \\[-0.3cm]
 date & PA & positions& average seeing(*)\\ [-0.3cm] \\\hline \\ [-0.3cm]
99/08/01&102\degr&0.6''S,0.3''S,0.0'', &0.5''\\
 & & 0.3''N,0.6''N&0.5''\\
99/09/17&12\degr&0.0'',0.3''W,0.6''W&0.45''\\
00/12/13&12\degr&0.3''E,0.6''E&0.75''\\
00/12/14&102\degr&0.9''S,1.2''S,1.5''S&0.9''\\
01/01/06&12\degr&0.9''E,1.2''E&0.37''\\
01/02/07&12\degr&1.5''E&0.55''\\
\hline\\
\label{obs_summary} 
\end{tabular}
\\
\end{center}
(*) Seeing values provided in this table have been taken from the ISAAC 
instrumental log. We have independently 
checked these values on the processed images wherever the slit was crossing 
through the nuclear unresolved K
continuum source. 
\end{table}
The slit width was set to 0.3'' while its length was of 2'. The slit was 
successively 
positioned at PA= 102\degr~
and PA=12\degr. The orientation PA=102\degr~corresponds to 
the orientation of the trace of the torus suggested by the near-IR 
observations of Rouan \etal~(\cite{rou98}), where we consequently expect 
to detect with maximum amplitude the velocity signature 
of a rotation. At the perpendicular orientation, PA=12\degr, no rotational
signature is to be expected. The slit positions were chosen to cover a 
spatially continuous area around the nucleus. 
Figure~\ref{ref_positions} shows the slit positions, superimposed over the 
AO M band image from Marco \& Alloin~(\cite{mar00}). 

The spectral resolution of the MR mode is 8700 which, at 2.15\micro, 
translates into a 35\kms~resolution. The projected scale on the detector is 0.147'' per pixel. Working 
in the IR, we used a standard 
nodding technique (408 pixels throw). The on-source individual exposure time 
is 270 seconds. Total 
integration times are in the range from
2$\times$270 seconds to 6$\times$270 seconds. The signal-to-noise ratio of the 
continuum reaches 100. 
Standard stars were acquired through 0.3'' and 2.0'' wide slits in order to 
respectively correct for sky absorption features and calibrate the data in 
flux.\\
\subsection{Data reduction}

Successive reduction steps were applied using the \texttt{IRAF} package, 
except for the electrical ghost 
removal, for which \texttt{Eclipse} was used.
Details on IR spectra reduction and especially on the nodding technique 
(differential comparison A - B and B - A) can be found in the ISAAC data 
reduction manual, Cuby \etal~(\cite{cub00a}). The main steps are the 
following : 

(1) correction of the electrical ghost: using the \texttt{Eclipse} is\_ghost 
routine, 

(2) simultaneous dark, bias and sky emission line subtraction: following the 
standard nodding technique,  

(3) flat-field division: carried out after correction of the illumination 
variation along the slit, 

(4) correction of 2D distortions (slit curvature and imperfect parallelism of 
the dispersion direction with the detector 
rows) using \texttt{IRAF},

(5) correction of emission line residuals (see nodding technique),

(6) extraction of the spectra,   

(7) wavelength calibration using the night-sky emission lines. The wavelengths 
are taken from the night-sky spectral atlas 
of OH emission lines in the near-IR by Rousselot \etal~(\cite{rou00}),

(8) correction of the sky absorption lines: using spectra of standard stars,

(9) flux calibration performed using stellar models found in the ``stellar 
spectra library'' (http://www.eso.org/instruments/isaac/lib/index.html).

\medskip

As we were interested in the spatial distribution of the line emission, we 
extracted a series of 1D-spectra from each 
2D-spectrum: 3 pixels-high i.e.
(0.45'' along the slit height) with a sliding step of 1 pixel
(0.147'').  The final spectral resolution is 2.5\,\AA, and the wavelength 
calibration sharp enough to measure the 
position of the night-sky lines within a precision of 0.1\,\AA. Over the 
central 0.3''$\times$0.3'' area, the continuum is quite
intense and produces a fringing pattern at the level of 5\% (peak-to-peak) 
which cannot be fully corrected. This is the region where the \H2~line is 
weak or absent. 
\\
\section{Data: Measurements and results}

\subsection{The \H2~and \Bg~emission: intensity maps}
\begin{figure*}
\resizebox{18cm}{!}{\includegraphics*[scale=1.]{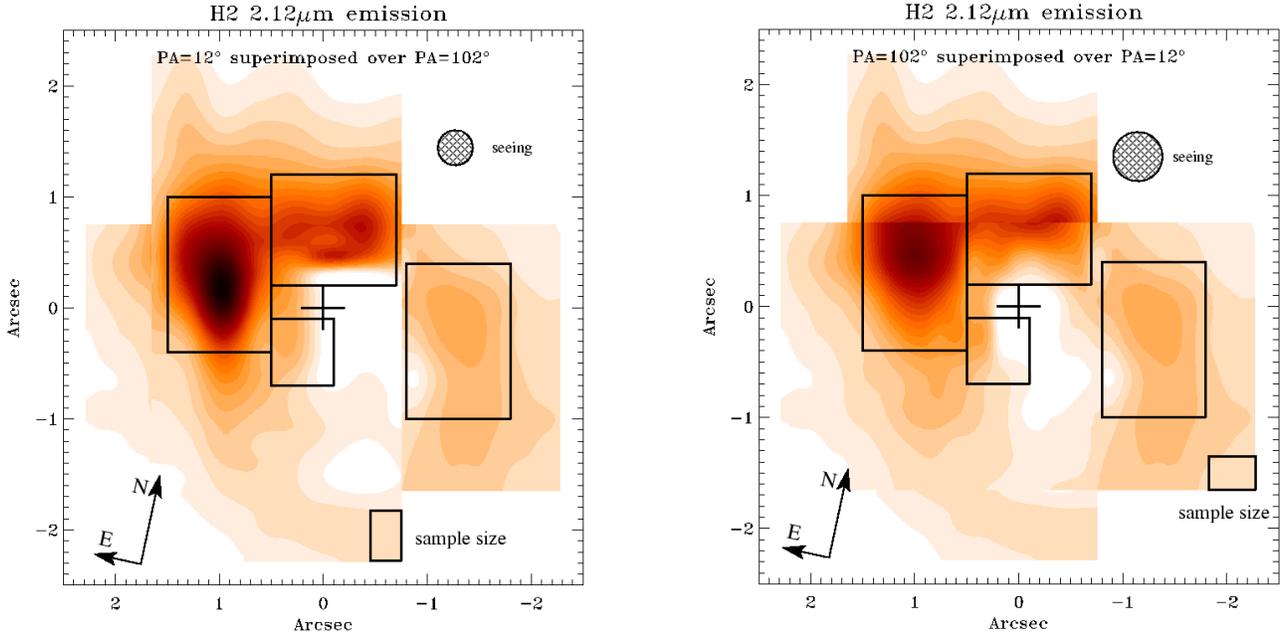}}
\hfill
\caption{Comparison between the H$_{2}$ maps reconstructed using the independent data sets at PA=102\degr~and PA=12\degr. 
The seeing value given on each figure corresponds to the seeing value while the slit was crossing the intense eastern \H2~emitting 
knot. The maps have been rebinned to a smaller square-pixel to allow easier comparison between maps which 
have originally rectangular and differently oriented pixels. Similar structures appear on both maps. The impact of a 
better image quality on the sharpening of structures is also obvious. We show 
as well on these maps the spatial areas 
over which the total \H2, and if applicable \Bg, line fluxes have been 
measured and given in Table 2. These are, starting from the box to the East, 
and turning clock-wise: 
the \east-\H2~knot, the \north-\H2-\Bg~knot, the West-\H2~knot and the 
South-\Bg~knot (see section \ref{sec_4_1} for a detailed discussion). The 
location of the central engine is shown by a cross.}
\label{comparison map}
\end{figure*}

In each 1D-spectrum we measured the
total velocity-integrated fluxes in the \H2~and \Bg~lines above the continuum.
These measurements allowed us to reconstruct 2D
maps of these line emissions. Figure~\ref{comparison map} compares two
reconstructed maps in \H2, obtained independently from the data sets at both 
PAs.
On the left panel, the map reconstructed from the data set at PA=12\degr~has 
been
overlaid on top of the map reconstructed from the data set at PA=102\degr. 
The
seeing value indicated relates to the PA=12\degr~spectrum through the intense
\H2~knot to the East of the central engine (indicated by a cross). On the right panel, the map
reconstructed from the data set at PA=102\degr~has been overlaid on top of the
map reconstructed from the data set at PA=12\degr, and the indicated seeing is
for the PA=102\degr~spectrum through the \H2~knot to the East.
The consistency of results derived from independent spectral data sets and the
accuracy of the map reconstruction technique
are successfully tested through the similarity of the structures appearing on
both maps. The impact of a better image quality
is also obvious in the left panel of Figure~\ref{comparison map} where the
strongly emitting East-\H2~knot appears more sharply, as the seeing value 
was then down to 0.37''. Four particular regions are highlighted
on Figure 2, identified and discussed in section 4.1. These are the 
\east-\H2~knot, the \north-\H2-\Bg~knot, the West-\H2~knot and the 
South-\Bg~knot, denominated according to their major line contribution. 

The ISAAC \H2~map can be compared to the early
map presented in Rotaciuc \etal~(\cite{rot91}), within the limitation of
different
spatial resolutions. Considering only their \H2~image obtained under the
best
seeing conditions (Figure 2 in their paper), we find indeed that their results
are consistent with the ISAAC reconstructed maps, both in terms of positioning
of the \H2~knots identified and in terms of their relative intensity.
On the contrary, the \H2~line map presented by Blietz \etal~(\cite{bli94})
is very different from that by Rotaciuc \etal~(\cite{rot91}) and from the
current ISAAC data set. While almost no \H2~line is found at the location of
the near-IR continuum core, either by Rotaciuc \etal~(\cite{rot91}) or from
the ISAAC data, this is the strongest \H2~emission knot found by in
Blietz \etal~(\cite{bli94}) -- in addition to the other \H2~knots which are
in reasonable agreement with those detected by Blietz \etal~(\cite{bli94}) and in
the ISAAC data set. How could we explain this discrepancy? As the near-IR continuum core is very strong
indeed, an extremely careful continuum subtraction is mandatory if
one uses narrow-band imaging. The observed difference in \H2~line emission at 
the position of the
continuum core, could easily be understood
if the continuum had not been subtracted properly in the Fabry-Perot data set
by Blietz \etal~(\cite{bli94}), leaving a residual as a fake \H2 line
emission.\\
Before discussing the ISAAC maps in detail, it is important to assess the
level of the uncertainties in positioning and flux measurements.

\subsection{Estimation of the error-bars in positioning and flux measurements}

As mentioned above, we extracted from the 2D-spectra series of 1D-spectra through 0.45''(3 pixels along
the slit) $\times$ 0.3''(slit width) windows, with a sliding step of one pixel
(0.15'') along the slit. Each 1D-spectrum is assigned to a pseudo pixel on the
reconstructed image. Hence, the pseudo pixel size is 0.3''(slit
width)$\times$0.15''(sliding step along the slit), that is three times smaller
than the extraction window. This is equivalent to applying a median filter to an
image made of independent pixels. Each 2D-spectrum enables to reconstruct an 0.3''
wide strip of the final reconstructed image. These independent strips have to be
carefully positioned along the direction of the slit with respect to a common
reference position. Along each slit, the reference is set to the spectrum with
maximum near-IR continuum level. We have checked that this positioning procedure
works fine, using an acquisition image of the continuum in K. For cuts along
either PA=102\degr~or
PA=12\degr, we find that the near-IR continuum maxima are located within an 0.3''
box around the intersection of the cut direction and its perpendicular passing
through the core. This indicates that the relative positioning of the strips with
respect to the core is precise within better than 0.3''.
Moreover, as the telescope offsets are set with a precision greater than
$\pm$0.05'' and as the slit is 0.3'' wide, in
the worst case, any identified structure on the reconstructed map has its
coordinate perpendicular to the slit
known better than within an 0.3'' window. In conclusion, any feature on the
reconstructed images can be positioned with
respect to the core within a box 0.3''$\times$0.3''.

Regarding flux uncertainties, let us recall that the line fluxes are measured
from the reconstructed line maps.
To derive these maps we have combined spectra obtained under slightly different
image quality: this introduces an
uncertainty in the flux measurement, especially when measuring line
fluxes from small scale
structures on the image or from features with steep spatial intensity gradients.
Actually, this effect is minimized here for the following reason: we have measured the velocity-integrated
\H2~line flux in the East-\H2~knot, where the intensity gradient is steep, from a set of
combined spectra all obtained under
extremely good seeing (0.37''). In the other identified knots with \H2~line
emission, West-\H2~knot and North-\H2-\Bg~knot, the
intensity gradient is not as steep and then the
effect of slight image quality differences not important. On the other hand, the great advantages of
our technique over using direct images acquired through
filters centered on the lines of interest are first, that the continuum subtraction
is very precise and second, that we avoid any loss of line emission resulting from
an unknown and unexpected
velocity field of the emitting material, since the line profile is accessible. Therefore, the final
precision on flux measurements depends on the flux calibration step itself which,
in this wavelength domain and given the very good observing conditions,
is expected to be better than $10\%$.

\subsection{Line flux measurements and comparison with previous measurements\\}

We provide in Table~\ref{tab_flux} the measured line fluxes in \H2~and \Bg~of the structures identified in Figure~\ref{comparison map}, \east -\H2~knot, 
West-\H2~knot, \north -\H2-\Bg~knot and South-\Bg~knot. The ``apertures'' used 
to measure the 
fluxes are shown by the boxes sketched on Figure~\ref{comparison map}. It 
should be noted that the exposure time indicated in the headers of files 
related to the ISAAC observations of July and August 1999 are incorrect. 
Consequently, the flux measurement in the \H2~line published by Alloin 
\etal~(\cite{all01}) is higher than the real value by a factor 6. The correct 
flux values are those given in the present paper. This error in the headers 
of ISAAC data files has been corrected for data taken after March 2000, but 
remains so far in the headers for data prior to this date. This will be 
modified in the future. Considering the flux calibration 
uncertainty ($\pm$$10\%$) and the possible slight flux losses due to seeing 
effects and estimated to be less than $5\%$, we get a final flux precision 
of around $\pm$$10\%$.
 
For comparison with previously published measurements, those are recalled 
in Table~\ref{pub_flux}. The dispersion in the \H2~and \Bg~line fluxes 
measured by various groups is large and there appears to be a monotonic decrease with time. At least two reasons can be invoked to 
explain the spread: (a) aperture differences and (b) intrinsic variability 
of the AGN in \NGC1068. Regarding the second point, it is well established that the near-IR 
continuum luminosity of the AGN in \NGC1068 has been increasing steadily 
since 1974 (Marco \& Alloin 
\cite{mar00}, Glass \cite{gla97}). 
However, notice that the \H2~line emission is arising from regions located at 
least 40\pc~away from the central engine (\north -\H2-\Bg~knot): these 
regions are today in a state of excitation corresponding to the level of 
activity of the central engine about 100 years ago. Conversely, the near-IR 
continuum luminosity is known to arise from a core less than 0.12'' (8\pc) 
according to Rouan \etal~(\cite{rou98}) or even as small as 0.03'' (0.2\pc) 
according to speckle measurements by Wittkowski \etal~(\cite{wit98}): it may 
have increased over the past 25 years, without impacting yet on the \H2~line 
emission. This may explain the observed decoupled behaviour between the
near-IR continuum flux and \H2~line time variations. Would well-sampled 
light-curves be available in the near-IR continuum and \H2~line, then 
time-delay effects could be used indeed 
to probe in another way the size of the \H2~line emitting region 
(e.g. Peterson~\cite{pet94}).\\
We have measured as well the continuum value at 
2.18\micro~on the ISAAC spectra passing through the core. Taking into 
account the measured seeing (FWHM of a cut through the continuum), 
considering two values for the intrinsic core size of respectively 0.12'' 
and 0.2'' FWHM, and given the 0.3'' slit width, we have computed the 
corresponding slit losses. The final values for the flux of the core 
continuum at 2.18\micro~are respectively 2.06\,$10^{-10}$ and 2.16\,
$10^{-10}$\densflux, for an 0.12'' and 0.2'' FWHM core. The value derived by 
Rouan \etal~(\cite{rou98}) 
from AO observations of the unresolved core in K was of 0.8\,
$10^{-10}$\densflux.  
\begin{table}[h]
\caption[]{Line fluxes for the identified knots} 
\begin{center}
\begin{tabular}{c|ccc} \hline \\[-0.3cm]
Source&Aperture&Aperture&\H2~line flux\\
 & center & size &  \\ \hline \\[-0.3cm]
 & & &\\
\east -\H2&+1.0'',+0.3''&1.0''$\times$1.4''&2.3$\pm$0.3\\
West-\H2&-1.3'',-0.3''&1.0''$\times$1.4''&0.7$\pm$0.1\\
South-\H2&0.2'',-0.4''&0.8''$\times$0.8''&$0.15 \pm 0.05$\\
\north -\H2-\Bg&-0.1'',+0.7''&1.2''$\times$1.0''&1.3$\pm$0.2\\ 
Total&0'',0''&4.0''$\times$4.0''&6.1$\pm$1.0\\  
 & & &\\ \hline \\[-0.3cm]
Source&Aperture&Aperture&\Bg~line flux \\
 & center & size &  \\ \hline \\[-0.3cm]
& & & \\
\east -\H2&+1.0'',+0.3''&1.0''$\times$1.4''&$0.25 \pm 0.15$\\
West-\H2&-1.3'',-0.3''&1.0''$\times$1.4''&$<0.2$\\
South-\H2&0.2'',-0.4''&0.8''$\times$0.8''&$0.4 \pm 0.1$\\
\north -\H2-\Bg&$-0.1'',+0.7''$&$1.2''\times 1.0''$&$2.8 \pm 0.5$ \\
Total&0'',0''&4.0''$\times$4.0''&5.2$\pm$1.0\\  
 & & & \\
\hline
\end{tabular} 
\label{tab_flux}
\end{center}
Note: The flux unit is ${\rm 10^{-14}}$\flux. 
The apertures are shown on Figure~\ref{comparison map}~; the aperture centers are 
defined with reference to the position of the K band unresolved core, coincident -- within error-bars -- with the radio source S1 (coordinates J2000 RA=02 42 40.7098, Dec= - 00 00 47.938 after Bland-Hawthorn \etal~\cite{bla97}). Be aware that the uncertainty on this positional reference is around 0.1''.  
\end{table}

\begin{table}[h]
\caption[]{Previously published fluxes in the \H2~and \Bg~lines}
\begin{center}
\begin{tabular}{c|ccc} \hline\\[-0.3cm]
Author &  Aperture & \H2~flux & \Bg~flux \\  \hline \\[-0.3cm]
Hall (1981) & 3.8'' diam. & $16.5 $ & $18.5 $  \\
Oliva (1990) & 6'' diam. & $15. \pm 1$ & $13.\pm 2$ \\
Rotaciuc (1991) &$\approx$ 3'' diam. & $8.6 $ & $4. $ \\   
this paper (2002) & total$\approx$ 3'' diam. & $6.1 \pm 1.0$ & $ 5.2 \pm 0.5$ \\
\hline
\end{tabular} 
\label{pub_flux}
\end{center}
The flux unit is ${\rm 10^{-14}}$\flux. If not mentioned, the error-bars
were not provided by the authors.
\end{table}
 
\begin{figure*}
\resizebox{18cm}{!}{\includegraphics*[scale=1.]{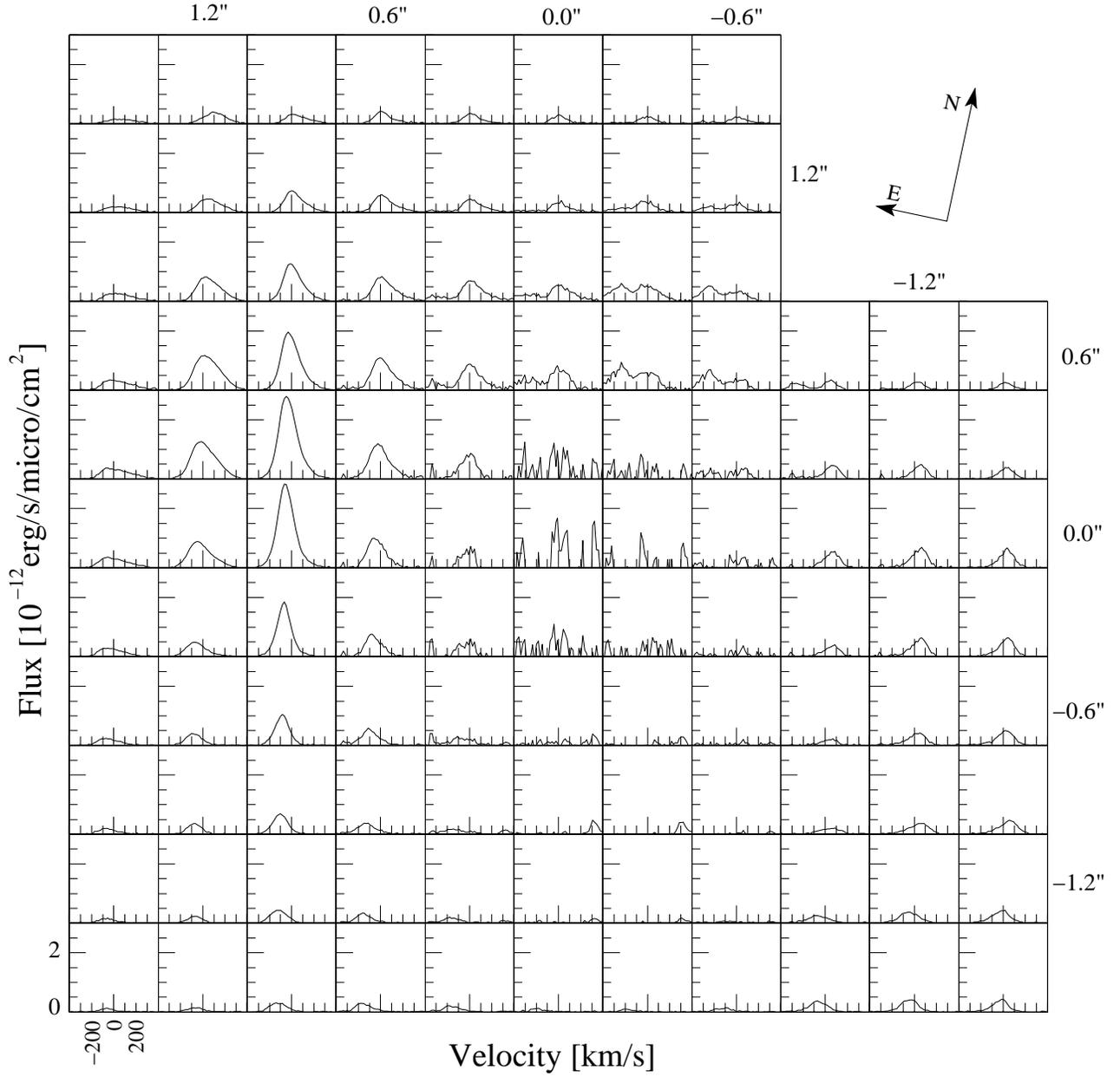}}
\hfill
\caption{The H$_{2}$ emission line profiles across the 3''$\times$3'' region around the nucleus. \north~is up, \east~is left. Each box corresponds to an emitting region of 0.45''$\times$0.3''. The unresolved continuum core in the K band is located at (0.0'',0.0''). The X-axis corresponds to PA=102\degr~and the Y-axis to PA=12\degr. The velocity scale of the profile is shown on the bottom left of the X-axis. The intensity scale is displayed on the bottom left of the Y-axis.}
\label{gridH2}
\end{figure*}
\begin{figure*}
\resizebox{18cm}{!}{\includegraphics*[scale=1.]{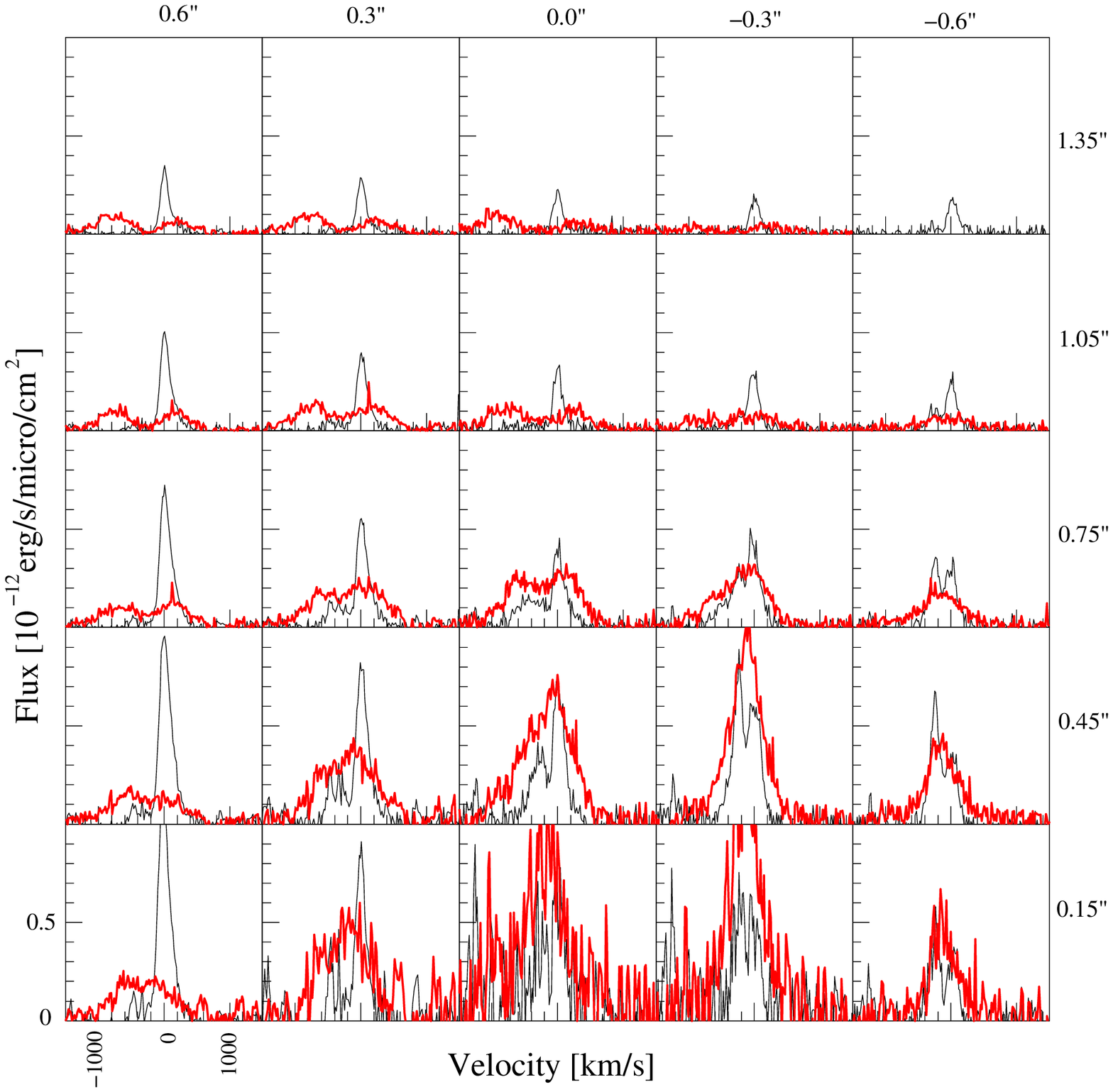}}
\hfill
\caption{The \Bg~and \H2~emission line profiles across the 1.2''$\times$1.2'' region to the North of the nucleus, corresponding to the North-\H2-\Bg~knot. The \Bg~line profiles are shown with a red and thick line while the \H2~profiles are shown with a black and thin line. Note that the X-axis velocity scale had to be changed because of much wider profiles in \Bg~than in \H2. As in figure~\ref{gridH2}, the X-axis corresponds to PA=102\degr~and the Y-axis to PA=12\degr. The orientation is the same as in figure~\ref{gridH2}.}
\label{gridBg}
\end{figure*}

\subsection{The \H2~and \Bg~emission: line profiles}
Figure~\ref{gridH2} and Figure~\ref{gridBg} provide respectively grids of 
the 
\H2~and \Bg~line profiles. In each box across this grid, we sample the line profile and flux from an 
0.3''$\times$0.45'' sub-area.  The X-axis provides its offset along PA=102\degr~(close to \east -\west) and 
the Y-axis its offset along PA=12\degr (close to \north -\south). Positive coordinates 
along both axes are respectively to the \east~and \north . The unresolved continuum core in the K band is referenced at 
coordinates (0.0'',0.0''). Regarding the central -0.3'' to +0.3'' strip, the 
strong near-IR continuum from the core induces a residual fringing level (3\%) 
and a noise level which prevent measuring weak \H2~line emission. Given the 
mean seeing value (0.5'' in the K band), line profiles in contiguous boxes 
are not fully independent one from the other. As the goal of this study is 
to understand and model the overall 
kinematics of the molecular gas across the studied region rather than its 
very local behaviour on a scale smaller than 0.5'', this 
inter-dependence does not limit much the interpretation.   
   

\begin{center}
\begin{figure*}
\resizebox{16cm}{!}{\includegraphics*[scale=1.]{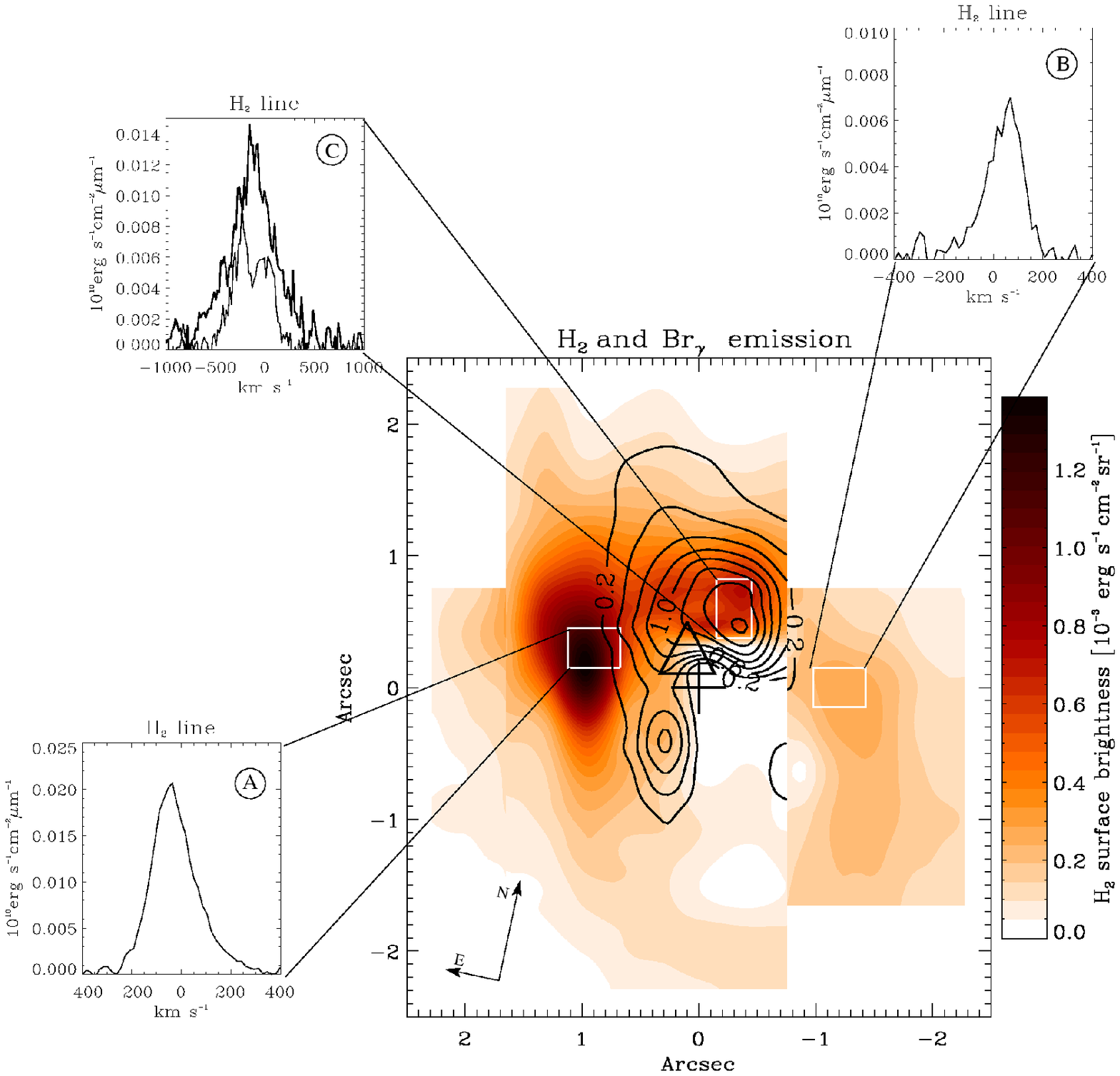}}
\caption{Molecular and ionized gas structures in the central 4''$\times$4'' 
of \NGC1068, as shown on a composite image made from the independent maps 
restored 
from the ISAAC data sets obtained 
at PA=102\degr~and PA=12\degr. The grey-scale map corresponds to the 
velocity-integrated \H2~line emission. We have superimposed 
with black contours the \Bg~line emission which is detected mainly to the 
North and South of the central engine. The labels on the contours are the 
surface brightness in $10^{-3}$\fluxsr. The cross
shows the location of the central engine (position of the unresolved core 
seen in the K band continuum: its coincidence with
the radio source S1 is ascertained within the error-bar of AO measurements 
by Marco \etal~\cite{mar97}). The triangle indicates the 
location of the radio source C. In the inserts A, B and C, we have displayed 
the \H2~line profiles (thick line) and the
\Bg~line profile (thin line) for three particular regions, the \east-\H2~knot, 
the \north-\H2-\Bg~knot and the West-\H2~knot. The 
profiles are displayed on a velocity scale to allow an easy comparison: 
the velocity zero point corresponds to the rest wavelength of each of 
the lines after correction for the heliocentric recession velocity of 
the galaxy, 1144 $\pm$4\kms, measured on our data as the mean velocity of the 
two sides of the suspected torus (see section 4.1). 
}

\label{fig_main}
\end{figure*}
\end{center}

\section{Kinematics and distribution of the emitting \H2 material
and of the ionized hydrogen over the central 4''$\times$4'' 
area}

\subsection{Emitting \H2~material and ionized hydrogen over the central 4''$\times$4'' area}
\label{sec_4_1}
Figure~\ref{fig_main} summarizes some of the conspicuous facts to be 
concluded from these observations: superimposed over the map
in the \H2~emission line, the contours indicate the flux in the 
\Bg~emission line, while the cross features the location of the K band 
continuum maximum on our 
ISAAC data set.  
It has been shown by Marco \etal~(\cite{mar97}) and Rouan \etal~(\cite{rou98}) 
from AO images in the K band, that this unresolved core (less than 0.12\pc~in 
size) is coincident, within error-bars, with the radio source S1 (central 
engine). The triangle shows the 
location of the radio source C (Muxlow \etal~\cite{mux96}; Gallimore 
\etal~\cite{gal96}). Marco \etal~(\cite{mar97}) also showed that the radio 
source C coincides with the [OIII] emitting cloud NLR-B seen on the HST maps.

We observe that the \H2~line emission is not distributed uniformly 
around the central engine. It consists of a low level, extended and diffuse 
component which is minimum where the continuum is maximum (central near-IR 
core at (0.0'',0.0'')). In addition to this extended \H2~line emission, three 
knots which are particularly intense in the \H2~line emission have been identified: the
corresponding line profiles (\H2~and/or \Bg) are displayed in the inserts 
in 
Figure~\ref{fig_main}. The most conspicuous one is the \east -\H2~knot, well 
defined in position and velocity (insert A). We can isolate as well the 
West-\H2~knot (insert B), although it appears weaker in intensity (factor 
around 3). No \Bg~emission is detected at the location of the West-\H2~knot, 
while the East-\H2~knot does show a low level of \Bg~emission. This emission 
can be attributed to the wings of the strong \Bg~emission coming from the 
North-\H2-\Bg~knot, rather than being intrinsic to the East-\H2-knot itself. 
We notice that, 
within  positional uncertainties, the \east -\H2~knot and
the West-\H2~knot are located almost symmetrically with respect to the 
position 
of the central engine. Their \H2~line peak velocities are respectively at 
1074$\pm$4\kms~and 1214$\pm$4\kms. Under the assumption they are kinematically 
associated in the form of a structure rotating around the central engine, this leads to a rest 
velocity of the central engine of 1144$\pm$4\kms. This velocity is fully 
consistent with the HI heliocentric velocity of the host galaxy, 1148$\pm$5\kms 
(Brinks \etal~\cite{bri97}). In addition, the \H2~line profiles in the East-\H2~knot and West-\H2~knot have similar 
width and are symmetrical in shape relative to the galaxy rest velocity. 
The line profile of the \east -\H2~knot has a blue-shifted peak and shows 
an extended red 
wing, i.e. toward lower velocities. The line profile of the West-\H2~knot has a 
red-shifted peak and also displays a wing toward lower velocities (extended 
blue wing). The \H2~line profiles (shape and velocity) change continuously 
and smoothly all over the region of diffuse emission and the identified 
\east -\H2~and  
West-\H2~knots: hence we can reasonably assume that they belong to a same 
kinematical component. The third knot identified is located to the 
\north~(insert C). 
It shows different characteristics than the East-\H2~knot and West-\H2~knot 
and emits both in \H2~and 
\Bg. In this \north -\H2-\Bg~knot, the \H2~line profile is double-peaked. If
one considers only the red peak of the \H2~line in the \north -\H2-\Bg~knot, 
one finds that the \H2~line profile peak velocity 
varies in an organized fashion from the eastern side to the western side 
of the central 
engine. Conversely, in the 
\north -\H2-\Bg~knot, the blue-peak of the \H2~line profile is unique to this
location, which is spatially coincident with the strongest \Bg~line emission 
detected. Although 
at this location the 
\Bg~line does not show an obvious double-peaked profile in correspondence 
to that observed in the \H2~line, one cannot exclude that it might be the possible 
blend of two 
components with greater FWHM ($\sim$700\kms) than the \H2~line components 
($\sim$190\kms). Indeed the \Bg~line 
emission is expected to arise from the NLR where the 
presence of ionized material at high velocity is inferred from the observation 
of broad wings in the [OIII] line profiles (Alloin \etal~\cite{all83}, Cecil 
\etal~\cite{cec90}). Let us remind finally the South-\Bg~knot which appears 
to be a stronger emitter in \Bg~than \H2. Notice that the \Bg~emitting knots,
the \north -\H2-\Bg~knot and South-\Bg~knot, are also quasi-symmetrical with 
respect to the central engine and outline the two sides of the ionizing cone.\\
From our results we can infer the following structure of the molecular 
and ionized material:
\begin{itemize}
\item The molecular \H2~line emission from (a) the diffuse component, (b) 
the \east -\H2~knot, (c) the West-\H2~knot and (d) the red peak on the 
\H2~profile from the \north -\H2-\Bg~knot, all arise from a same system with 
organized kinematical properties. This is demonstrated by the continuous 
spatial change of 
the \H2~line profile peak velocity. Within this system, the maximum velocity 
shift observed between the \east -\H2~knot and the West-\H2~knot 
has a value of 140\kms, these knots being located at a distance 
of $\simeq$1'' (70\pc) on each side 
of the central engine. The simplest interpretation of this velocity 
jump is in terms of the signature of a rotating structure: under the 
assumption of keplerian rotation, this implies a central mass of 
$10^8$\,M$_{\odot}$ (Alloin \etal~\cite{all01}). 
Modeling of the kinematics of this system, derived from
the \H2~line profile changes over the entire region, is presented 
in Section 4. 
\\
\item The \Bg~line emission and the blue-peak of the \H2~line emission 
in the \north -\H2-\Bg~knot, as well as the \Bg~emission in the South-\Bg~knot
belong to a kinematically distinct system. This system is probably 
associated 
with the NLR material, as suggested by the location of the \north -\H2-\Bg~knot 
and South-\Bg~knot, by the absence of velocity offset of their line profile
peaks and by the large width of the 
\Bg~line. In addition, let us notice that the \north -\H2-\Bg~knot is 
found to be close to the radio 
source C identified on the MERLIN map (Muxlow \etal~\cite{mux96}). 
The source C coincides with the re-direction of the small-scale radio 
jet (initial direction PA=12\degr), possibly induced by the impact on 
the NLR-B cloud (for details see Bland-Hawthorn \etal~\cite{bla97}). The 
ionization resulting from the impact would explain 
the detection of the [OIII] and \Bg~lines.        
\end{itemize}

\subsection{The molecular material over the 20''$\times$20'' area}
\begin{figure}[h]
\mbox{\resizebox{9cm}{!}{\includegraphics[scale=1.]{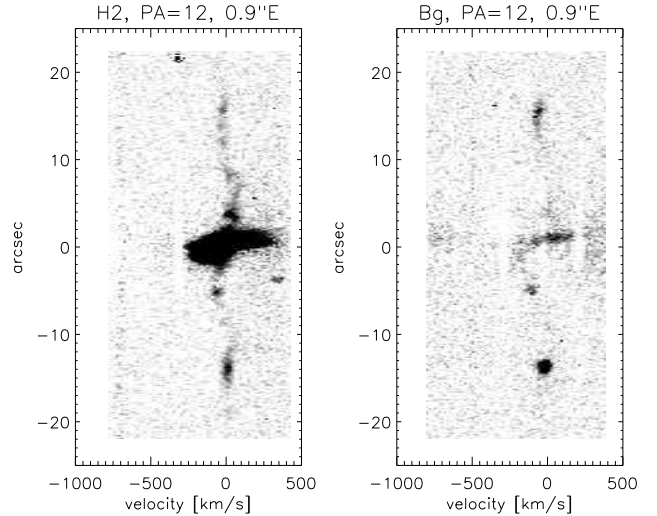}}}
\caption{Extended \H2~and \Bg~emission up to 15'' from the nucleus and with the slit passing through the East \H2 knot. The left and right figures display respectively the (P-V) diagram for \H2~and for \Bg, along PA=12\degr.}
\label{fig_extend}
\end{figure}

The ISAAC long slit data set also provides information about the distribution of the \H2 line emission over a larger scale. The 2D spectra, particularly 
along PA=12\degr, show weak \H2~and \Bg~emission up to almost 20'' from the 
nucleus on both sides. These two lines are most clearly detected between 14'' 
and 18'' from the nucleus to the North, and between 12'' to 17'' from the 
nucleus to the \south . The locations of these various emission spots correspond to 
the star forming spiral arms clearly traced by the CO millimeter lines 
(Tacconi \etal~\cite{tac97}, Schinnerer \etal~\cite{sch00}). The line 
intensity ratio \Bg/\H2~is about 3 to the \north~and 1 to the \south . 
These lines have a FWHM of 35\kms~and are much narrower than the emission lines in the central 
4''$\times$4'' region. The 
spatial coverage of our data set is too scarce to extract complete 2D 
information over the central 40''$\times$40'' where the CO arms and bar 
develop. Still, we notice that, along PA=12\degr~the presence of \H2~and 
\Bg~is conspicuous (Figure~\ref{fig_extend}), while these lines are 
rather weak along PA=102\degr. Along PA=12\degr, the 
velocity curve does not exhibit a simple behaviour. Along PA=102\degr~the velocity curve 
traced by \H2~and \Bg~is consistent with a logarithmic potential: zero 
on the nucleus and then rising up to a plateau.         

\section{Kinematics: a rotating disk and an outflow}

An important constraint on the kinematics of the molecular material 
around the central 
engine in NGC1068 comes from the velocity jump of 
140\kms~between the East-\H2~and the West-\H2~knots. An ad-hoc jet 
configuration along PA-102\degr~might produce such projected velocities; 
however, as the ionization 
cone axis and the radio emission all appear to be aligned 
along PA=10\degr, it seems rather natural to interpret the observed molecular 
emission distribution and kinematics in the framework of a somewhat 
axisymmetrical structure located in a plane perpendicular 
to the axis defined at PA=10\degr. 
Another strong constraint on the kinematics comes from the asymmetry of 
the observed \H2~line profiles. 
They exhibit a more extended wing on the side of their small velocities: 
ie. blue 
wing for the profiles which are redshifted with respect to the galaxy 
rest-frame 
velocity, and red wing for the profiles which are blueshifted. This 
observational fact must be accounted for.
\subsection{Modeling the line profiles: a model with two kinematical components}

\begin{figure}[h]
\mbox{\resizebox{9cm}{!}{\includegraphics[scale=1.]{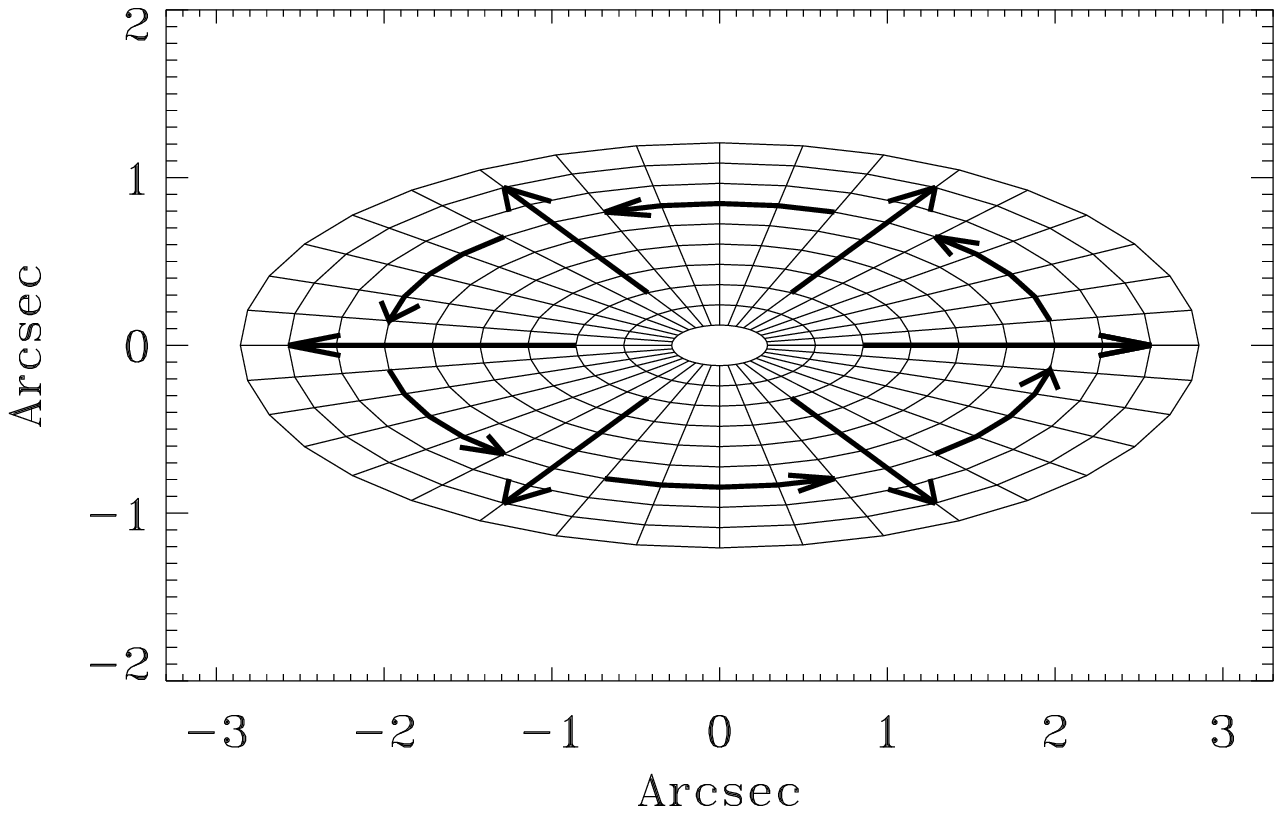}}}
\caption{Illustration of the two-component kinematical model. Arrows represent the motion of the two components.}
\label{disk}
\end{figure}

The hypothesis of a rotating structure for the molecular material is 
tested through a simple model. The model is used to produce a grid of 
\H2~emission line profiles to be compared to the observed ones over the 
4''$\times$4'' observed region. The model allows to produce profile grids 
for either 
of two independent kinematical components and for their combination. The 
two kinematical components considered are: a rotating disk and an outflow, 
as illustrated on Figure \ref{disk}. 

For the rotation of the disk we assume a velocity field induced by a 
logarithmic 
gravitational potential. Such a velocity field is described by two 
parameters: a characteristic radius $r_p$ and a characteristic velocity $v_p$. 
The velocity 
curve flattens at radii greater than $r_p$, and reaches a velocity $v_p$. 
The outflow is only parameterized through the outflow velocity. 
\begin{center}
\begin{figure*}[t]
\mbox{\resizebox{9cm}{!}{\includegraphics[scale=1.]{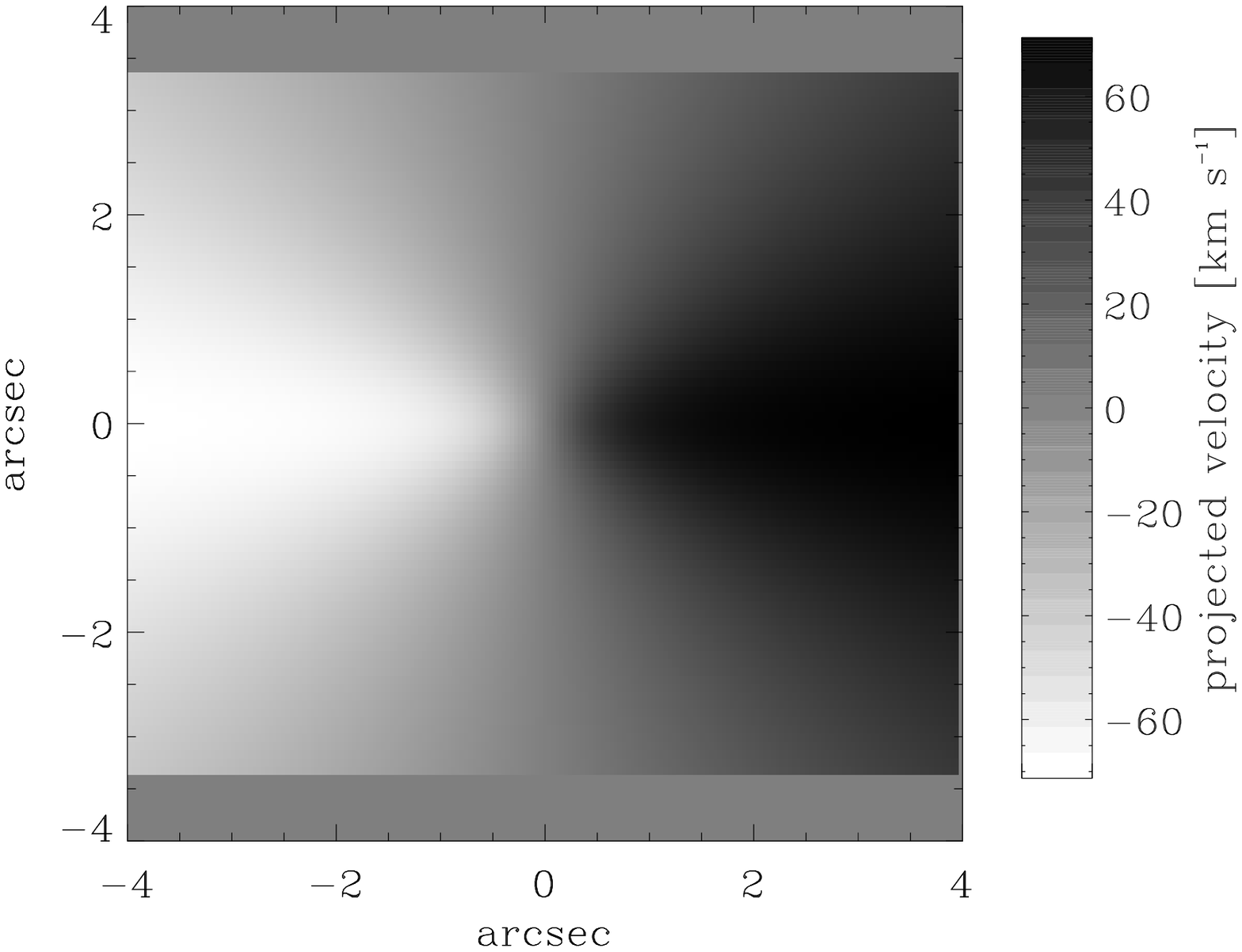}}}
\mbox{\resizebox{9cm}{!}{\includegraphics[scale=1.]{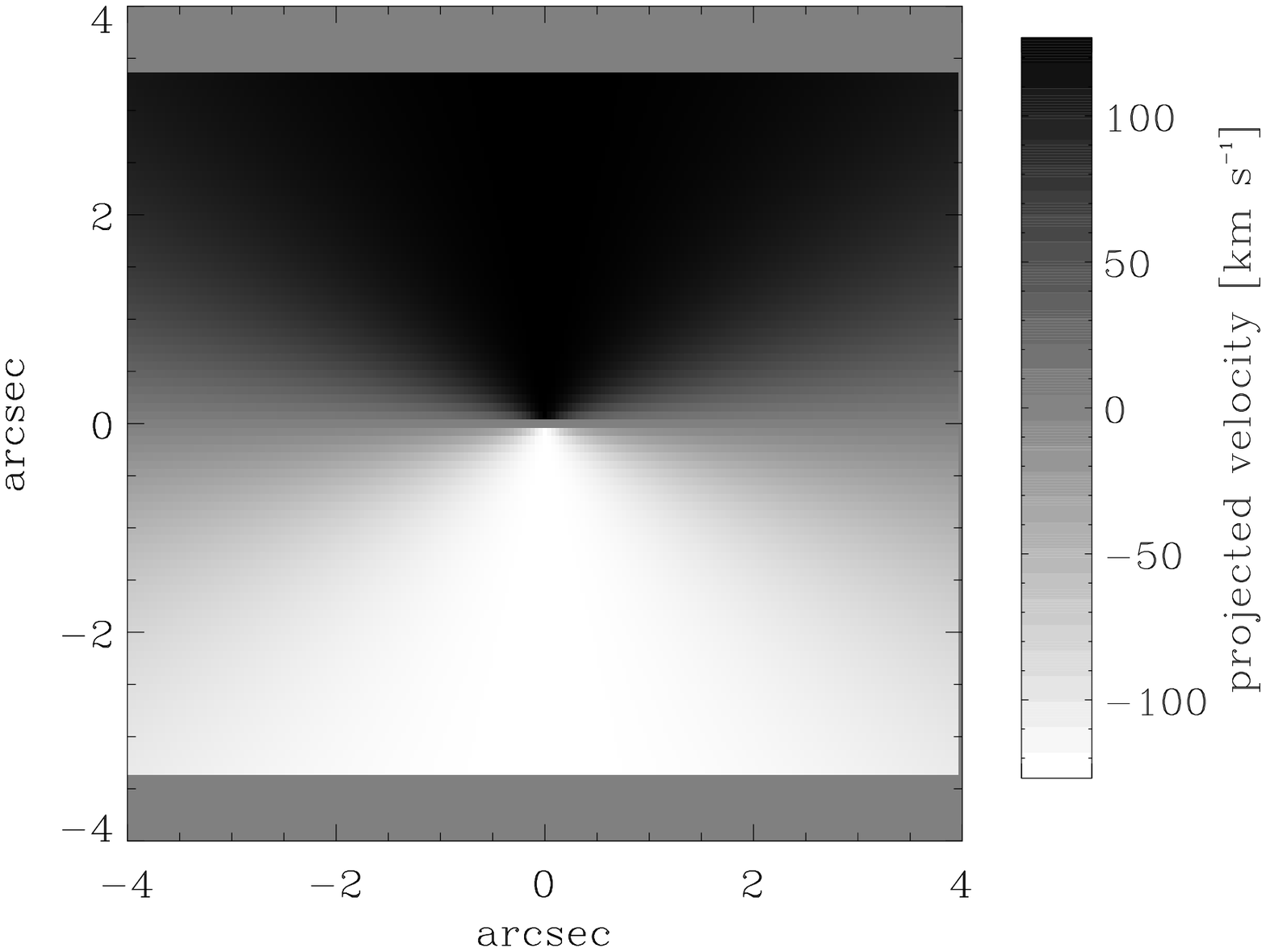}}}

\caption{Along-the-line-of-sight $v_{z}$ velocity map of the rotating 
component (left) and of the outflow component (right) in the considered model}
\label{fig_im_vit_wind}
\end{figure*}  
\end{center}

\begin{figure*}[t]
\resizebox{18cm}{!}{\includegraphics*[scale=1.]{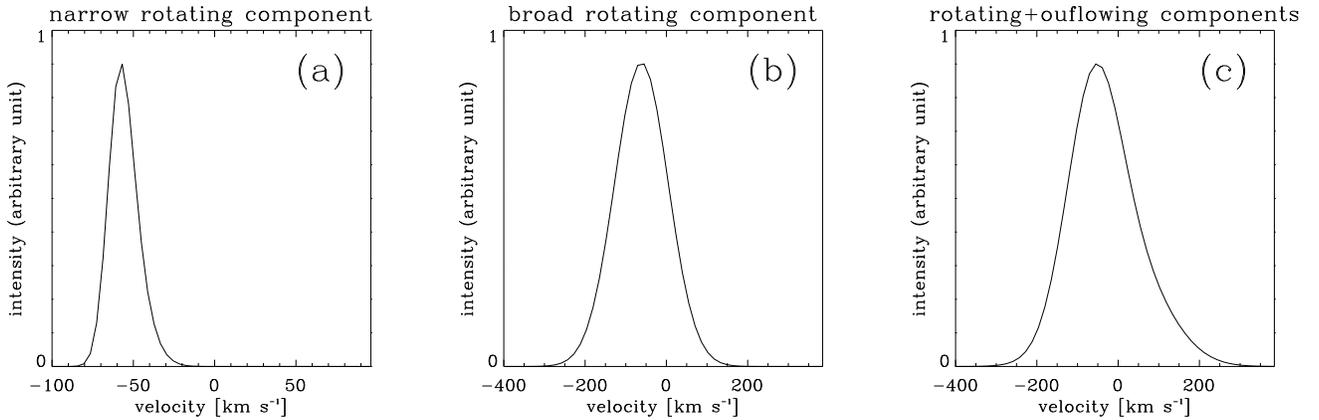}}
\hfill
\caption{Simulated profiles for the position [1''E,0''N] on the grid. From 
left to right: (a) rotating component alone, intrinsic line FWHM=15\kms~and 
instrumental profile not taken into account: the profile is asymmetric because 
of the mixture of different projected velocities resulting from the seeing. 
(b) rotating component alone, with intrinsic line FWHM=150\kms: the profile 
is symmetric because the intrinsic width is much greater than the range of 
projected velocities in a seeing sized-area. (c) rotating plus outflowing 
components with intrinsic line FWHM=150\kms~for the rotating component and 
line FWHM =200\kms~for the outflowing component: this combination allows to 
reproduce at the same time the width and asymmetry of the observed profiles, 
as well as the spatial evolution of the asymmetry.}
\label{three_mod_prof}
\end{figure*}
The model is built in the following way: a 2D grid, where each element 
represents a small surface of the disk, is placed in the plane of the sky. 
We choose the X-axis to represent PA=102\degr~and the Y-axis to represent 
PA=12\degr. 3D vectors are assigned to each element of the grid: one for 
the position and one for the velocity of each kinematical component. A 
rotation around the X-axis, to account for the inclination of the disk, is 
applied to these vectors. Then a $v_{z}$ ``image'' is built for each 
kinematical component; we assign to each pixel of this image the 
along-the-line-of-sight velocity of the grid element whose projected position 
on the sky corresponds to the pixel position. The $v_{z}$ images for a 
rotating and for an outflowing component are shown in 
Figure~\ref{fig_im_vit_wind}. From each $v_{z}$ image, we build a 3D 
cube where the third dimension is the velocity. A Gaussian profile is 
used to represent the emission line and is centered at the corresponding 
$v_{z}$ value. The intensity assigned to the profiles was chosen to be constant, 
after checking that this assumption does not affect our conclusions. 
Each plane from this data cube is then convolved with a 2D Gaussian 
representing the point spread function (PSF) to account for seeing effects. 
The parameters for the kinematical model are the following:
\begin{itemize}
\item the velocity field parameter(s) (two parameters for the rotation and one 
for the outflow)
\item the inclination of the disk
\item the intrinsic width of the emission line for each kinematical component
\item the relative intensity of the emission from each of the kinematical components
\end{itemize}

The \H2~line profiles  
predicted in the two configurations -- with or without outflow -- are shown 
in Figure~\ref{three_mod_prof}. We find that it is not possible to reproduce 
{\bf at the same time} the general characteristics 
(asymmetry, width and peak position) of the observed \H2~line profiles using a 
model with 
only a 
rotating disk. On the contrary, the two-component model (rotating disk plus 
outflow) reproduces well all the profile characteristics and we shall pursue
the profile fitting with such a kinematical model. 
Do we expect, from a modeling point-of-view, such an outflow to be present 
in the environment of the AGN? In 
classical AGN models, the observed 
conical funneling of the ionizing radiation field is explained by inflation 
of the inner part of the molecular/dusty torus. One may therefore anticipate 
that the impact of the 
X and UV photons on the edges of the inflated inner regions of the torus 
will sweep away material from its surface and generate an 
outflow.\\

\subsection{Results}
\label{model}
\begin{figure*}
\resizebox{18cm}{!}{\includegraphics*[scale=1.]{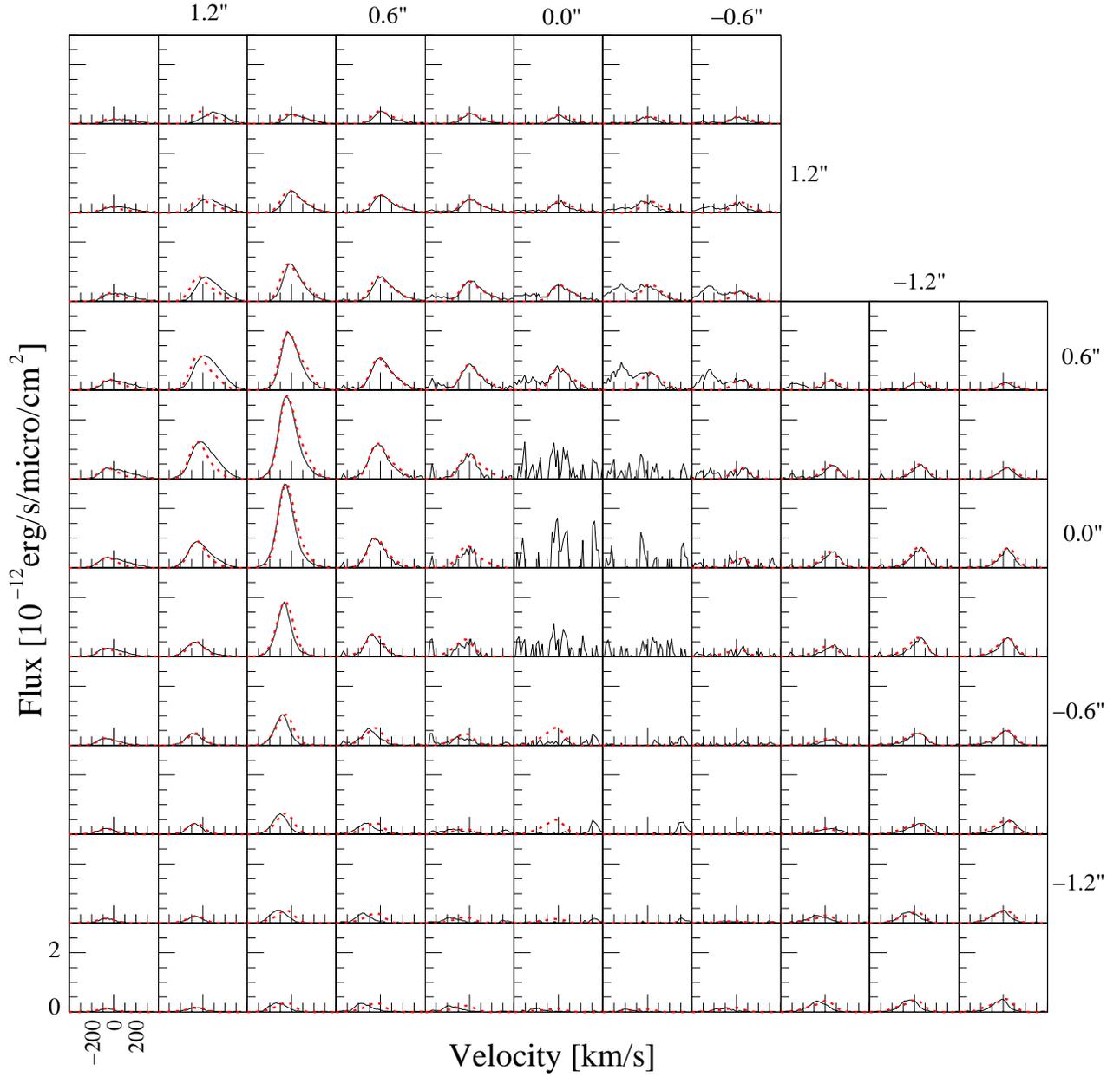}}
\hfill
\caption{Modeled profiles (dotted lines) and observed profiles (solid lines) over the 3'$\times$3'' central area. Values of the model parameters can be found in section~\ref{model}.} 
\label{model_result}
\end{figure*}
The simulated 
\H2~profiles over the 3''$\times$3'' region are plotted on top of the observed 
ones in the grid shown in 
Figure~\ref{model_result}. The parameter values used to build 
the grid of simulated profiles matching best the observed ones, are 
the following:    

\begin{itemize}
\item $v_p$= 100\kms
\item $r_p$= 50\pc
\item outflow velocity: 140\kms
\item inclination angle: 65\degr
\item FWHM of the rotating component: 150\kms~
\item FWHM of the outflowing component: 200\kms
\item ratio between the line intensities emitted by the rotating and outflowing component: 2
\end{itemize}
The modeled profiles have been scaled to peak intensity in order to allow 
an easier 
comparison among the profile shapes. The agreement between the observed 
and modeled \H2~line profiles is very good over the entire area. This 
validates the choice of 
a two-component kinematical model. One should note however that the figures 
derived  
for the model parameters are not constrained in a unique manner and
some variation around the quoted figures may be expected as line
profiles from other molecular transitions arising in the warm molecular
gas will become accessible to observation.\\ 

In comparison, the CO data set discussed by Schinnerer \etal~(\cite{sch00}) 
has been modeled in the framework of a warped disk. In the CO modeling, 
little emphasis 
has been put on fitting the CO line profiles, while in the case of the \H2~line 
emission, the profiles are used as strong constraints for the modeling. 
Let us recall in addition  
that low order CO transitions probe cold molecular material, while the 
\H2~transition studied here 
arises from warm molecular matter. Therefore a detailed comparison between 
the two types of models is 
hampered by the use of different constraints on one hand, and by reference 
to different temperature phases of the molecular material, on the other hand. 
We note however that the main
emission knots detected in \H2 are located at positions coincident with the
CO emitting knots. It 
is quite conceivable that the \H2~line originates at the surface of 
molecular clouds exposed to intense radiation fields, while the CO line 
arises from the denser and cooler inner part of these clouds. In such a 
configuration, surface-effects (wind...) would have a great impact on the 
shape of the 
\H2~line profile while not affecting the shape of the CO line profile.

\section{Discussion: source of excitation of the \H2~line}
Apart from the \north -\H2-\Bg~knot, the lack of strong \Bg~emission 
coincident with  
\H2~emission suggests that most of the molecular material emitting the 
\H2~line is not directly irradiated by UV photons which would also 
ionize hydrogen. Supporting a similar conclusion is 
the 90\degr~angle offset between the ionization cone axis (along the 
South-North 
direction, Macchetto \etal~\cite{mac94}) and the predominant direction of 
\H2~line emission (East-West) in \NGC1068. Moreover, it has been shown 
independently that the observed \H2 line ratio 2-1\,S(1)/1-0\,S(1) in 
\NGC1068 is consistent with thermal emission rather than with UV-pumping 
(Oliva and Moorwood \cite{oli90}). Therefore, these three facts lead us to 
disregard UV photons as the source of excitation of the \H2~line in the close
environment of the central engine in \NGC1068.\\
The remaining two possibilities to excite \H2~to the requested $\sim$2000\,K 
are, 
(a) heating by shock waves, or (b) heating due to irradiation by 
\Xrays~(Mouri \cite{mou94}).\\
Could shocks play a 
dominant role in the excitation of the \H2 emission in the particular case 
of \NGC1068? Following Draine, Roberge \& Dalgarno (\cite{dra83}), the shock 
velocities required are of the order of 
$V_{shock} \sim 10-30$\kms. An 
order of magnitude of the total shock luminosity is given by $L_{shock} 
\sim 1/2\,{\rho}_{cloud}\,V^3_{shock}\,A_{shock}$, where ${\rho}_{cloud}$ 
is the density of the cloud, $V_{shock}$ the shock velocity and $A_{shock}$ 
the total shock area (Heckman \etal~\cite{hec86}). 
A density $n_{\rm H_2}$=$10^5 \rm\,cm^{-3}$, a shock velocity 
$V_{shock}$=$30$\kms~and a shock area $A_{shock}$=$2{\pi}r_{cloud}^2$ for 
$r_{cloud}$=20\,pc are reasonable values for the East-\H2~knot. The total 
shock luminosity for these parameters is $\rm 10^{33}$\power. About one 
tenth of this luminosity is radiated through the 1-0\,S(1) \H2 line. The 
resultant predicted luminosity in this line is thus 6 orders of magnitude 
below the observed $\rm 3.10^{38}$\power~emitted in the East-\H2~knot. Even 
if we increase generously $n_{\rm H_2}$ or $A_{shock}$, the observed flux 
values cannot be accounted for in this way. Therefore we can disregard
shock heating as the source of excitation of the \H2~line around the
central engine in \NGC1068.\\
Maloney (\cite{mal97}) has already discussed the \Xray~irradiation 
of molecular clouds by the central \Xray~source in \NGC1068. 
His model was computed for an intrinsic 1-100\,keV luminosity of $10^{44}$\power~for the central engine and an attenuating column density of $10^{22}\,\rm cm^{-2}$ between the central engine and the \H2~emitting molecular clouds. With this set of parameters, his model accounts for the observed \H2~intensity and predicts the right size-scale of the \H2~line emission. However, Matt \etal~(\cite{mat97}) have shown that the opaque material which obscures our direct view to the central engine of \NGC1068~is Compton thick ($\rm N_{H} \geq 10^{24}\,cm^{-2}$). 
If this opaque material has a more or less toroidal distribution -- as suggested by the existence of the ionizing cone --, then the column density between the central engine and the molecular material found at the location of the East and West-\H2~knots must also be of the order of at least $\rm10^{24}\,cm^{-2}$. In such a configuration, the model predicts an \H2~emission arising much closer to the nucleus than the observed 70\pc. 
An alternative possibility is that the azimuthal density distribution in the torus is such that the opening angle for the \Xrays~is larger than for the UV radiation. This situation is likely to occur since \Xrays~can penetrate much larger column densities than UV photons, implying that the viewing angle range over which the torus is Compton thick is smaller than the viewing angle range over which it is UV-opaque. Still, in such a situation, the \H2~emitting clouds must be located within aperture of the cone inside which the \Xrays~can escape. This suggests that the molecular clouds located at a distance of 70\pc~(where the \H2~emission is mainly detected) are not 
strictly orbiting in the plane 
perpendicular 
to the axis of the opaque torus. Consequently, either the \H2~emitting 
material is a component distinct from the torus and is simply orbiting in 
the plane of the galaxy (inclination $\sim$ 40\degr), or we are dealing 
with a single disk structure within which the inclination of the plane, where 
the 
molecular material orbits, changes continuously with radius: in other words 
the disk is warped.\\
The presence of a warped structure in the AGN of \NGC1068 has also been proposed 
by Backer (\cite{bac00}) to fit both the intensities and the kinematics of 
the CO emitting material. Interestingly, as shown in a theoretical work 
by Quillen (\cite{qui01}), a wind blowing across a dense, thin and initially 
flat rotating disk would automatically generate a warp. This suggests that 
there is a possible physical connexion between the warp and the outflow, hence 
between the observed distribution of \H2~emission and the \H2~line profile 
shapes. 
First, the asymmetry of the profiles witnesses the presence of an outflow. 
Second, modeling predictions tell us that this wind would have naturally 
induced a warp of the molecular disk. Third, it is the existence of the warp 
which makes possible the irradiation 
of the molecular material by the central \Xrays. In summary, this scenario  
makes the wind the indirect cause of the \H2~emission being  
observed at a rather large distance from the central engine.      

\section{Concluding remarks}
We have provided in this paper a comprehensive study of the kinematics
of the warm molecular material in the environment of the AGN in 
\NGC1068. New runs of 2D spectroscopic observations (ISAAC on VLT/ANTU),
obtained under very good seeing conditions, have complemented the 
data set discussed earlier by Alloin \etal~(\cite{all01}).
The achieved spectral (35\kms) and spatial (0.5'') resolutions allow to 
recover emission maps in the \H2~and \Bg~emission lines, at respective 
rest wavelengths 2.122 and 2.166\micro~, over the central 4''$\times$4'' 
region surrounding the central engine in \NGC1068. In addition, one witnesses
the spatial evolution of the \H2~and \Bg~line profiles over the inner
3''$\times$3'' region with a spatial sampling of 0.3''$\times$0.3''.\\
The \H2~and \Bg~line emission maps and profile spatial evolution have been used to
analyze the distribution and kinematics of the warm molecular gas across the 
200\pc~central region. The salient results of this work are the following:
\begin{itemize}
\item
There is no detectable \H2~emission at the location of the strong 2.2\micro~continuum 
core (which is known already to be almost coincident with the central engine), 
while two main regions of \H2~emission, East-\H2~and West-\H2, are detected at 
about 70\pc~on each side of the central engine along PA=90\degr, with a 
projected 
velocity difference of 140\kms: this velocity jump is interpreted at first 
order as the signature of a rotating structure. Under this assumption, the 
systemic velocity of the AGN is found to be 1144$\pm$4\kms. 
No strong \Bg~emission is detected at the location of these knots. 
The East-\H2~knot is about three times brighter in \H2~line emission than the 
West-\H2~knot
\item
One detects as well a diffuse and extended \H2~emission, although at 
low intensity level. At about 0.3'' to the North of the central engine, the 
\H2~emission also coincides with a \Bg~line emitting knot. This 
\north -\H2-\Bg~knot is close to the radio knot C and the brightest 
[OIII] cloud, NLR-B, although not exactly spatially coincident with them.
\item
To the South, another \Bg~line emitting region is observed, although weaker
in intensity, the South-\Bg~knot.
The \north -\H2-\Bg~knot and South-\Bg~knot are almost symmetrical with respect
to the central engine and appear to be located on the inner edges of the 
ionizing cone.

\item
The observed changes in the \H2~line profile across the 200\pc~central region 
cannot be reproduced by the simple, disk-like, kinematical structure which 
matches the 
velocity jump between of the East-\H2~knot and the West-\H2~knot. A possible solution is 
to add an outflow at the surface of the rotating disk-like 
structure. The main parameters of the complete kinematical model are: a rotational 
velocity around 100\kms~at a characteristic radius of 50\pc~, an inclination 
angle of the disk-like structure of 65\degr~and an outflow velocity of 140\kms.
\item
The source of excitation of the \H2~line is briefly discussed. Both
UV photons and shocks can be discarded as the main source of excitation of the
\H2~line, while X-ray irradiation from the central engine is found to be 
the most likely mechanism.
\item 
An additional consequence of an outflow at the surface of a rotating
disk-like structure is the occurrence of a warp. This warp does naturally
explain how the X-rays can impact the molecular disk at a distance as
large as 70\pc~and give rise to the two observed main knots emitting \H2.    
\end{itemize}

\acknowledgements
We are gratefully indebted to the ESO Service Observing team on Paranal, 
especially to Jean-Gabriel Cuby and Olivier Marco who performed the 
service mode observations, and to the User Support Group and Archive Support 
Group at ESO/Garching for efficient help. We acknowledge precious advice 
from Chris Lidman for data reduction and interesting discussions with 
Andrew Backer and Michael Burton. Finally, we thank the anonymous referee for 
stimulating remarks.



\end{document}